\documentclass[a4paper,11pt]{article}
\usepackage{pos}
\usepackage{graphicx}  
\usepackage{float}     
\usepackage{xcolor} 
\usepackage{subcaption} 

\title{Forward Spectator Detector for CBM}

\author*[a, b]{Radim Dvořák}
\onbehalf{for the CBM Collabroation}

\affiliation[a]{Czech Technical University in Prague,\\
  Jugoslávských partyzánů 1580/3, Prague, Czech Republic}

\affiliation[b]{Nuclear Physics Institute, The Czech Academy of Sciences,\\
25068 Rez, Czech Republic}

\emailAdd{dvorar10@fjfi.cvut.cz}

\abstract{The development of the Forward Spectator Detector (FSD) for the CBM experiment represents a crucial step toward successful realization of the CBM physics program - understanding of highly compressed nuclear matter at the forthcoming FAIR facility. Designed
for detecting collision participants at high collision rates at the SIS-100 accelerator, the FSD employs scintillator-based detector technology to accurately reconstruct the reaction plane and to determine the centrality of the collision.
Overview of the technical design and performance studies
conducted for the FSD is provided.}

\FullConference{FAIR next generation scientists - 8th Edition Workshop (FAIRness2024)\\
 23-27 September 2024\\
Donji Seget, Croatia\\}

\begin{document}

\note[]{This work has been supported by the Ministry of Education, Youth and Sports of the Czech Republic under the Large Research Infrastructures program MSMT LM2023060, and MSMT OP VVV CZ$.02.01.01\/00\/23\_015\/0008181$.}

\maketitle

\section{Forward Spectator Detector for CBM experiment}

\begin{figure}[h]
    \centering
    \begin{subfigure}{0.6\textwidth}
     \centering
     \includegraphics[width=\textwidth]{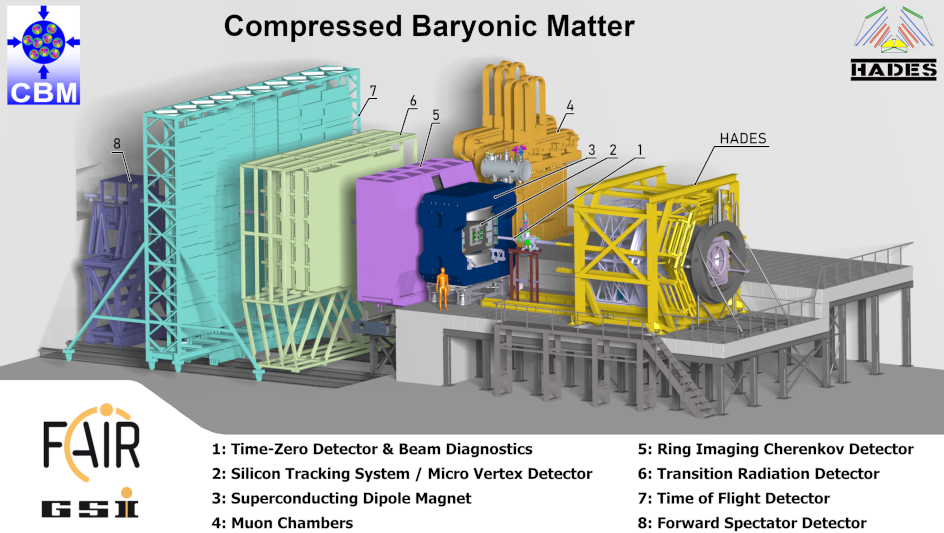}
     \caption{}
     \label{fig:cbm_model}
     \end{subfigure}
     \hfill
      \begin{subfigure}{0.38\textwidth}
    \centering
    \includegraphics[width=\textwidth]{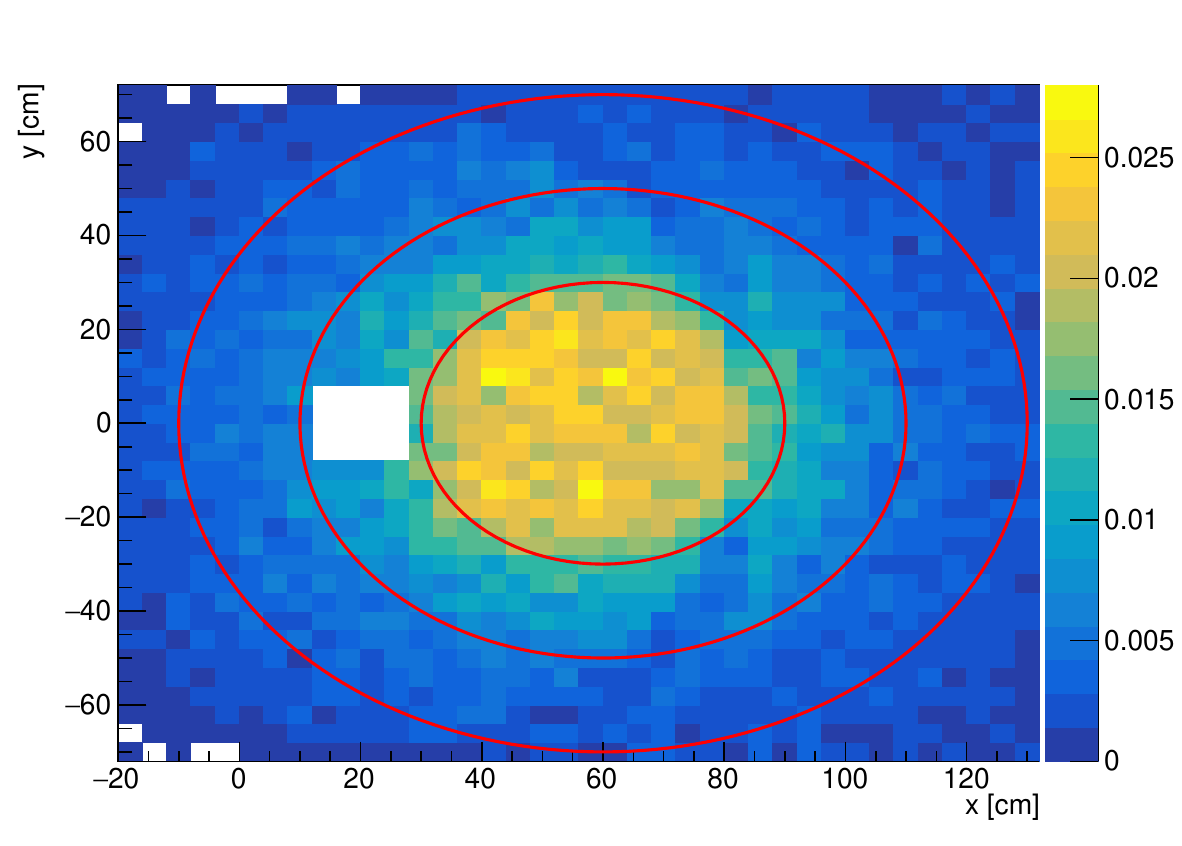}
    \caption{}
    \label{fig:proton_xy}
   \end{subfigure}
   \caption{\ref{fig:cbm_model} CBM experiment, \ref{fig:proton_xy} Distribution of hits from primary protons in rapidity $2.8<y<3.6$ in FSD. Red circle shows the subevents regions. IN: $r=[0,30]$ cm, MID: $r=[30,50]$ cm, OUT: $r=[50,70]$ cm.}
   \label{fig:1}
 \end{figure}

The Compressed Baryonic Matter (CBM) experiment \cite{CBM-Website} (see Figure \ref{fig:cbm_model}) is part of the future FAIR facility \cite{FAIR-BTR}. Its main aim is to study dense nuclear matter at high baryonic densities and exploration of the phase diagram of strongly interacting matter. The maximum beam kinetic energy for heavy ions will be $11A$GeV in Au+Au collisions. The collision rate will reach up to $10$ MHz, which will allow to study even very rare probes. The CBM is a fixed target experiment with a dipole magnet and two possible setups - optimized for electron or muon detection, respectively. In both configurations, the FSD, as the name suggests, is the most forward detector positioned close to the beampipe. The main aim of the FSD is to detect primary protons and fragments to measure the event-plane and centrality of the collisions. The current geometry of the FSD  consists of two layers at $14$ and $17$ m downstream from the target. Both detector planes are 150x140 cm in size and are composed of scintillator pads of 4x4 cm. The granularity and dimensions will be further optimized for the final design. The light from the scintillators is detected by photomultipliers and read out using DiRICH+TRB electronic, a well-proven technology developed by GSI \cite{TRB_GSI}. 

\section{Measurement of the flow}

Anisotropic transverse flow \cite{eliptic} is one of the most important observables for (ultra)-relativistic heavy-ion collisions. It describes the spatial anisotropy of the produced particles and can be expressed using the Fourier decomposition of the transverse spectrum:

\begin{equation}
\frac{dN}{d\phi} \propto  1 + 2 \sum_{n=1}^\infty v_n \cos \left[ n \left(\phi- \Psi_{RP} \right) \right], \quad v_n = \langle \cos \left[ n \left(\phi- \Psi_{RP} \right) \right] \rangle,
\label{rov:flow}
\end{equation}

where $\phi$ is the azimuthal angle of the particle, $v_n$ represents the $n$-th harmonic of anisotropic flow and $\Psi_{RP}$ is the reaction plane. The coefficient $v_1$, the so-called directed flow, tells if more particles are produced in the direction of the impact parameter or the opposite direction. The coefficient $v_2$ (elliptic flow) is an essential observable for studying the collective behavior of the matter produced in heavy-ion collisions. For obvious reasons, the reaction plane $\Psi_{RP}$ is not an observable quantity and has to be determined indirectly. Different procedures were developed to measure the flow and estimate the reaction plane. In our studies, the QnAnalysis framework is used \cite{QnAnalysis}. It is based on the q-vectors calculated as:

\begin{equation}
\mathbf{Q_n} = \frac{1}{M} \sum_{i}^N w_i \mathbf{u_{n,i}}, \mathbf{u_n} = \lbrace \cos n \phi, \sin n \phi \rbrace .
\label{rov:Qn}
\end{equation}

Here, $w_i$ denotes the weight of the i-th $\mathbf{u_{n, i}}$ vector and $M$ is the normalization factor. The q-vectors can be calculated for a variety of particles or hits in the detector. In our case, hits in FSD are used. The QnAnalysis framework was specifically developed to deal with the effects of nonuniform acceptance and efficiency corrections \cite{Selyuzhenkov2008}. 
The resolution of the event-plane obtained from the FSD is defined as:

\begin{equation}
 R_{x,y} = 2 \langle Q_{1 x,y} \Psi_{RP x,y} \rangle ,
\label{rov:resolution}
\end{equation}

where the $Q_1$ vector is calculated using Eq. \ref{rov:Qn} from hits in the FSD and the $\Psi_{RP}$ is taken directly from a Monte Carlo simulated event. Note that the resolution can be obtained separately for x and y directions. This is important since the dipole magnetic field created by the CBM magnet can have a significant influence on the resolution in x direction. However, the final results on $v_n$ should be independent of the direction used for the calculation. To be able to obtain the event-plane and its resolution in a fully data-driven way from hits in FSD only, the 3-subevent method is used \cite{Lubynets2024}. In this method, hits in FSD are divided into three subevents (A, B, C) and resolution is defined for each subevent based on the correlations between respective Q-vectors:

\begin{equation}
 R^A_{x,y} = \sqrt{2\frac{\langle Q_{A x,y} Q_{B x,y} \rangle \langle Q_{A x,y} Q_{C x,y} \rangle}{\langle Q_{B x,y} Q_{C x,y} \rangle}}
\label{rov:resolution_3sub}
\end{equation}

and similarly for two other subevents. One will hence obtain in total six independent values of resolutions using this method. The 3-subevent method is usually checked for autocorrelation between each subevent using factorization:

\begin{equation}
\langle Q_{A x,y} Q_{B x,y} \rangle = \langle Q_{A x,y} \Psi_{RP x,y} \rangle \langle Q_{B x,y} \Psi_{RP x,y} \rangle.
\label{rov:factorisation}
\end{equation}

If the correlations between subevents are not caused solely by the directed flow of participants the Eq. \ref{rov:resolution_3sub} will not be valid. To suppress these non-flow effects the 4-th subevent (D) can be introduced and resolution is then calculated using:

\begin{equation}
R^A_{x,y} = \frac{2 \langle Q^A_{x,y} Q^D_{x,y} \rangle}{R^D_{x,y}}, R^D_{x,y} = \sqrt{2\frac{\langle Q_{B x,y} Q_{D x,y} \rangle \langle Q_{C x,y} Q_{D x,y} \rangle}{\langle Q_{B x,y} Q_{C x,y} \rangle}}.
\label{rov:4sub}
\end{equation}
 The 4-th subevent is independent of the FSD. In our case, pions in forward rapidity ($2.4<y<2.8$) are used. The resolution for the two other subevents is calculated similarly by reshuffling the indexes. The calculated resolution is needed to obtain the flow:
 
 \begin{equation}
 v_{n x,y} = \frac{2 \langle q_{n x,y} Q_{n x,y}\rangle}{R_{x,y}},
 \label{rov:v1}
 \end{equation}
 
 where $q_n$ is a q-vector calculated from the measured particles (in our cause pions in $2.2<y<2.4$). The procedure then yields six values of $v_n$ which should be in principle the same but can be used to access systematic uncertainties due to detector nonuniformities.

\section{Performance studies}

The presented results were obtained using the DCM-QGSM-SMM model \cite{Botvina1995, Baznat2020} coupled with Geant4 \cite{Agostinelli2003}+CBM detector response simulator \cite{cbmroot}. The Figure \ref{fig:proton_xy} shows the distribution of the spectator protons of rapidity $y>2.8$ in the FSD layer at 17m from the target. As can be seen, due to the dipole magnetic field, the maximum of the proton distribution corresponding to beam rapidity of $y=3.2$ is around $x=60$ cm. The subevents used for the event-plane reconstruction are calculated from concentric regions around this position. Note that there is a non-active region in the detector at $x=20$ cm which corresponds to the place where beampipe passes through the FSD detector.

\begin{figure}[h]
    \centering
    \includegraphics[width=0.75\textwidth]{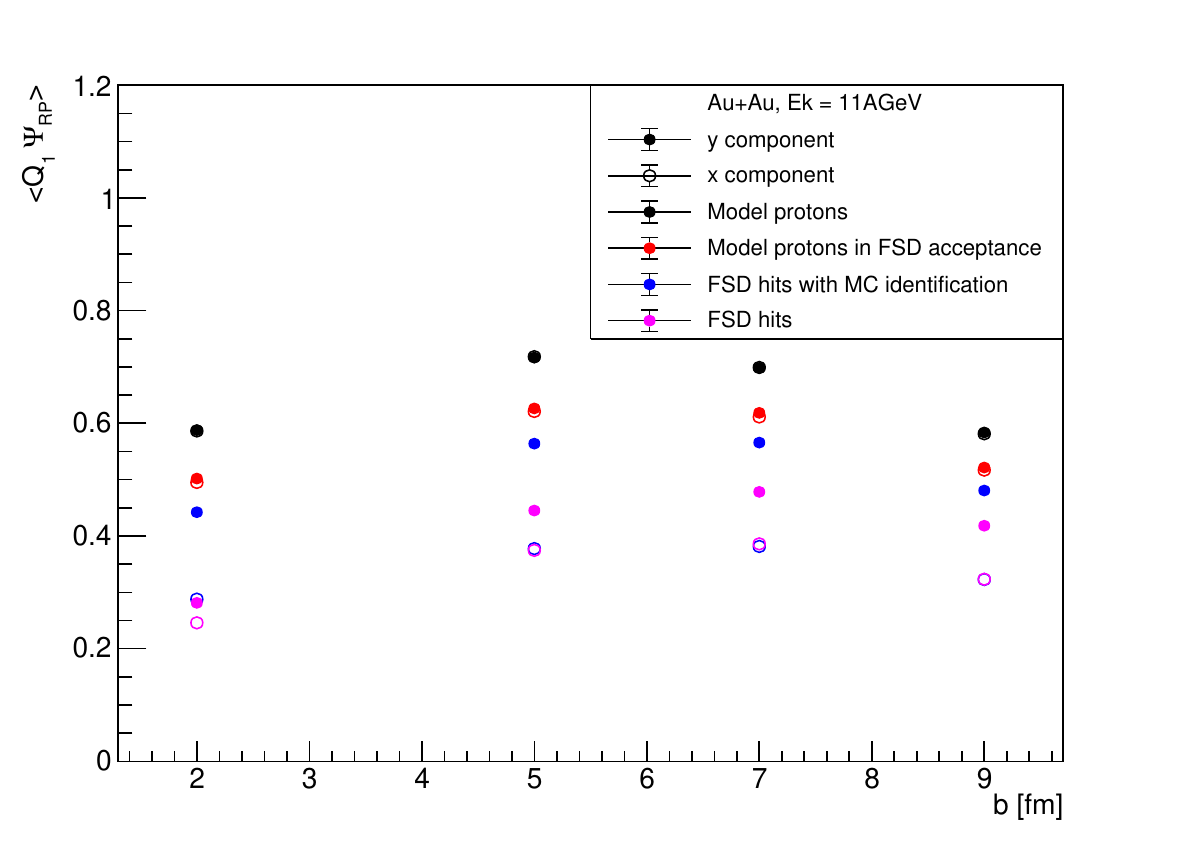}
    \caption{Comparison of event-plane resolution calculated from primary protons and FSD hits dependent on the impact parameter $b$.}
    \label{fig:resoltuion_comp}
\end{figure}

To assess the performance of the detector, the event-plane resolution obtained from the simulated FSD hits is compared to the one obtained directly from the model according to the theory framework described above. The comparison is shown in Figure \ref{fig:resoltuion_comp}. For all results there are separately shown values calculated from the x (open symbols) and y (full symbols) components of the q-vectors from Eq. \ref{rov:Qn}.  
 In Figure \ref{fig:resoltuion_comp} black points show the resolution calculated using Eq. \ref{rov:resolution} directly from the MC model. The maximum resolution is around $70\%$ for both components. The effect of the acceptance of the FSD detector is shown using red symbols. It is shown to be about $10\%$ lower, similar for both components. The nontrivial effect of the dipole magnetic field can be seen when one selects FSD hits which are, based on the MC information, known to belong to primary protons. Note that while the $y$ component is almost unaffected, the resolution in the $x$ direction drops significantly. In reality, the MC information about particles is not available and protons are identified using the deposited energy in the scintillator. Purity of the sample of primary protons is increased when coincidence between two planes is required such that the line connecting two hits points to the primary vertex. As can be seen, the expected resolution obtained purely from FSD will reach up to $45 \%$ in the $y$ component and $40 \%$ for the $x$ component for mid-peripheral events.

\begin{figure}[H]
      \centering
    \includegraphics[width=0.75\textwidth]{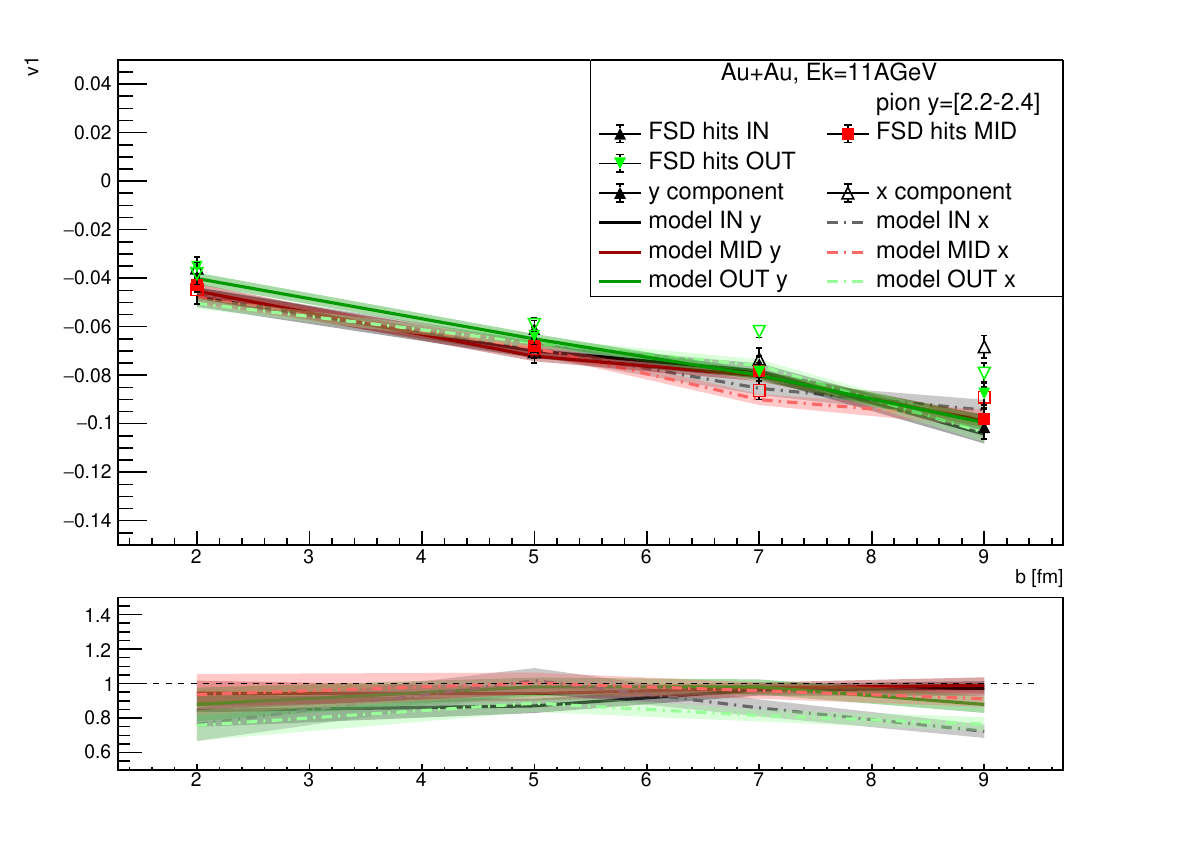}
    \caption{$v_1$ of pions in rapidity $2.2<y<2.4$ calculated using 4-subevent method. Symbols are calculated using FSD hits (y component - full symbols, x component - open symbols). The lines are obtained directly from the DCM-QGSM-SMM model.}
    \label{fig:matched_v1_4sub}
\end{figure}

The best crosscheck of the performance of the detector is to compare the final results on $v_1$ obtained directly from the MC event with those from the detector simulation using the same analysis technique. In Figure \ref{fig:matched_v1_4sub} results are shown on $v_1$ calculated using the 4-subevent method as described in the first section. The symbols are obtained using only simulated FSD hits for the reconstruction of the event-plane while the lines are coming directly from the MC model. As can be seen, it is possible to obtain the values of directed flow from the FSD detector with good precision in the $y$ component. The $x$ component shows a mild disagreement for peripheral events, caused likely by interaction of spectator fragments with beam pipe. This phenomenon is currently under investigation. Further studies on the optimization of detector performance such as granularity and final position are also under way.

\end{document}